\documentclass[12pt,preprint]{aastex}

\usepackage{natbib}

\def\kms{{\rm{km}}/{\rm{s}}}
\def\msun{\rm{M_{\odot}}}

\def\rhoc{\rho_{\rm{cld}}}
\def\rhopsf{\dot{\rho}_{\rm{sf}}}
\def\csfr{c_{\rm{sf}}}

\begin{document}

\title{Sub-Galactic Clumps at High Redshift: A Fragmentation Origin?}

\author{Andreas Immeli\altaffilmark{1}, Markus Samland\altaffilmark{1}, 
        Pieter Westera\altaffilmark{2}, Ortwin Gerhard\altaffilmark{1}}

\altaffiltext{1}{Astronomisches Institut der Universit{\"a}t Basel, 
                 Venusstrasse 7, CH-4102 Binningen, Switzerland}
\altaffiltext{2}{Observat{\'o}rio do Valongo, Universidade Federal 
                 do Rio de Janeiro, Ladeira do Pedro Ant{\^o}nio, 43, 
                 CEP 20080-090, Rio de Janeiro, Brazil}


\begin{abstract} 
  We investigate the origin of the clumpy structures observed at high
  redshift, like the chain galaxies.  We use a three dimensional
  chemodynamical simulation describing the dynamics of stars and a
  two-phase interstellar medium, as well as feedback processes from
  the stars.  For high efficiency of energy dissipation in the cold
  cloud medium, the initially gaseous disk fragments and develops
  several massive clumps of gas and stars.  We follow the evolution of
  the individual clumps and determine their masses, metallicities and
  velocities.  A few dynamical times after fragmentation of the disk,
  the clumps merge to build a massive bulge.  Calculating HST- and
  UBVRIJHKLM-colors, including absorption by interstellar dust, we
  determine the morphologies and colors of this model in HST images.
  Several peculiar morphological structures seen in the HDF can be
  well-explained by a fragmented galactic disk model, including chain
  galaxies and objects consisting of several nearby knots.
\end{abstract}

\keywords{Galaxies: evolution --- Galaxies: formation --- 
Galaxies: high-redshift --- Galaxies: structure --- Galaxies: peculiar}



\section{Introduction}

In the redshift range $0.5<z<3$, galaxies evidence a large diversity
of morphological types \citep[e.g.][ vdB96]{abraham01, steidel96,
  vandenbergh96}.  Although some of the unusual morphological
structures can be explained by the morphological K-correction, NICMOS
observations \citep{dickinson00} show that many galaxies indeed have a
rest frame optical morphology that cannot be attached to the
traditional Hubble scheme.

Examples are the so-called chain galaxies that show elongated knotty
structures \citep[ CHS95]{cowie95}. Different scenarios have been
proposed to explain these structures.  CHS95 suggested that chains lie
in the redshift range $0.5-3$ and have a mass comparable to that of a
present-day galaxy.  They speculate that these objects may be linear
arrangements in space where star formation, once turned on, triggers
further star formation along the line of maximum density.  In some
models the chains form in colliding supershells blown out of massive
starburst galaxies \citep{taniguchi01}.  \citet{dalcanton96} argued
that LSB galaxies are local counterparts to chains at high or
intermediate redshift.  \citet{oneil00} suggested, from a comparison
with less inclined objects, that the chains do not belong to a new
galaxy class but are knotty disk like structures seen edge on. This is
consistent with recent observations from
\citep{elmegreen04a,elmegreen04b}, who find that the clump colors in
face-on clumpy objects are similar to the colors of clumps in chain
galaxies.

Here we investigate a model of a gaseous disk that becomes unstable
and develops several clumps of gas and stars.  The model presented
here describes one of the evolutionary paths that a disk can take, in
the sequence investigated in \citet{immeli04}, with a higher-resolution
simulation.  The evolution of this disk is similar as in the scenario
proposed by \citet{noguchi98}, but our model for the star-forming
two-phase interstellar medium, including feedback processes, allows us
to keep track of stellar ages and metallicities, and thus to determine
realistic luminosities and colors for a direct comparison with
observations in the Hubble Deep Field (HDF). We show here that several of
the unusual morphological types in the HDF are well-described by a
fragmented disk model seen from different viewing angles.


\section{The Model}

We use a two-phase model for the interstellar medium, consisting of a
hot, low-density phase and a cold cloud medium from which stars are
formed. We describe this system with a three-dimensional
chemodynamical evolution code, which combines a hydrodynamical grid
code for the two phases of the interstellar medium (ISM) with a
particle mesh code for the stars.  The interactions between the
different ISM phases are described in \citet[ SG03]{samland03}. For
the star formation rate (SFR) we use a Schmidt Law \citep{schmidt59},
$\rhopsf = \csfr \cdot \rhoc^{\alpha}$ with $\alpha=1.5$ and $\csfr$
consistent with the star formation (SF) rule derived by
\citet{kennicutt98}.

The energy released from supernovae mostly goes into heating the hot
phase, but also heats the cold phase.  Most of the kinetic energy of
the cloud fluid is in the motions of single clouds relative to the
bulk flow.  This kinetic energy can be dissipated by inelastic
collisions \citep{larson69} and augmented by supernova feedback
\citep{mckee77}.  Its energy dissipation rate, here described by the
parameter $\eta_c$, is not well-determined, and may well vary between
galaxies. One expects that it depends on the geometrical structure of
the clouds, on whether a major part of the dense medium is arranged in
filaments, and on their self-gravitating structure and magnetic fields
\citep{kim01, balsara01}. The influence of $\eta_c$ on the dynamical
evolution of gas-rich disks is investigated in more detail in
\citet{immeli04}; together with the infall rate it determines the
amount of star-forming cold gas in the disk. Here we compare one of
their fragmenting disk models ($\eta_c=0.5$) with observations of
high-redshift galaxies.

The setup of our model describes an early and rapid formation of a
massive galactic disk in a static dark halo.  According to
\citet{sommer02} the delayed infall of the baryonic matter into the
relaxed halo can solve the angular momentum problem arising in
$\Lambda$CDM structure formation simulations.  The primordial gas
enters the simulation volume at $|z|=$ 7~kpc vertically and uniformly
distributed over a radius of 17~kpc, with a rotation velocity equal to
the circular velocity at the infall point, and infall velocity $20
\,\kms$.  The infall rate is $120~\msun yr^{-1}$ during one Gyr,
resulting in a total baryonic mass of $1.2 \cdot 10^{11}~\msun$.  The
simulation volume has a diameter of 37.2~kpc and a vertical height of
14~kpc with a spatial resolution of 300~pc in the horizontal and
120~pc in the vertical direction.  We have also done the simulation at
lower resolution, with similar results, indicating that the outcome is
not sensitive to the resolution used \citep{immeli04}.

        \begin{figure}
        \plotone{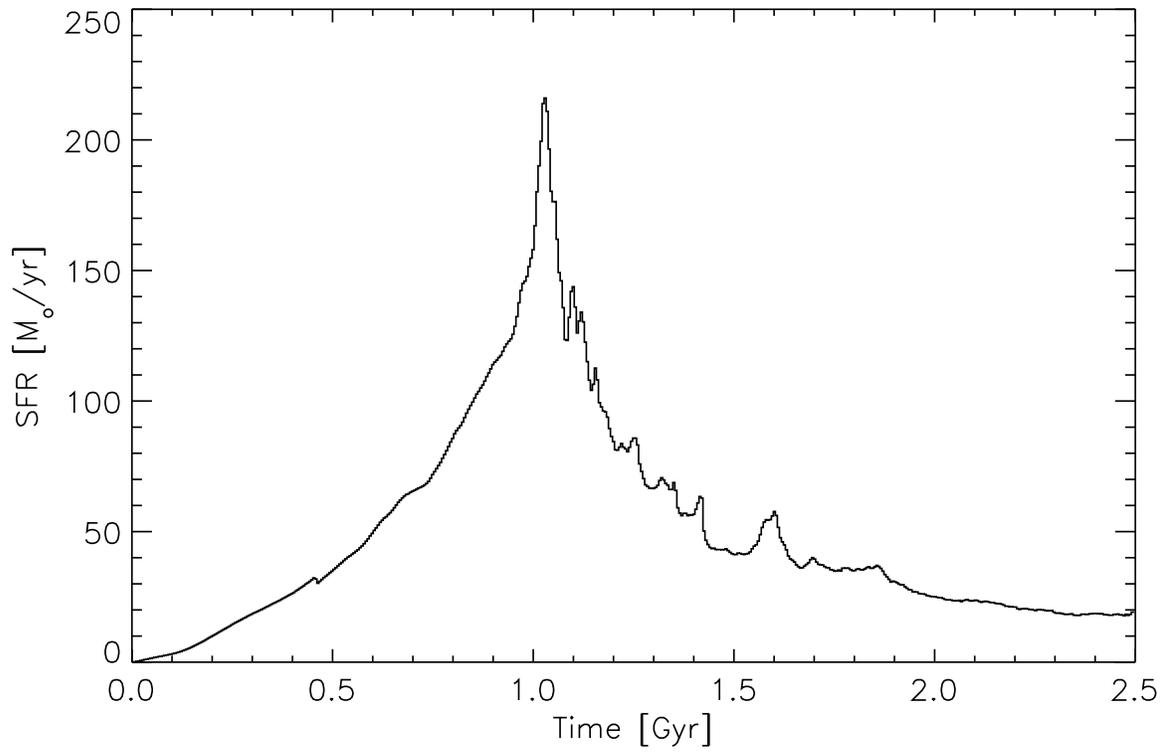}
        \caption{Star formation rate of the gas-rich disk model discussed
                 in this paper.}
        \label{sfrinf}        
        \end{figure}

The chemodynamical model provides ages and metallicities of the stars
formed, as well as ISM densities and metallicities. This enables us to
calculate colors of the model at different redshifts, including
absorption, using the method of \citet{westera02} except that we adopt
here a three times lower absorption coefficient.


\section{Results and Comparison to Observations}
\subsection{Global Evolution}
The infall of the baryons into the halo leads to the build-up of a
star-forming gaseous disk.  Fig.~\ref{sfrinf} shows the resulting
evolution of the star formation rate (SFR) in the model.  The energy
input from the supernovae type II dominates that from the infall
during most of the evolution and prevents the rapid formation of a
massive disk on a free fall timescale.

\begin{figure}
         \epsscale{0.7}                                                         
         \plotone{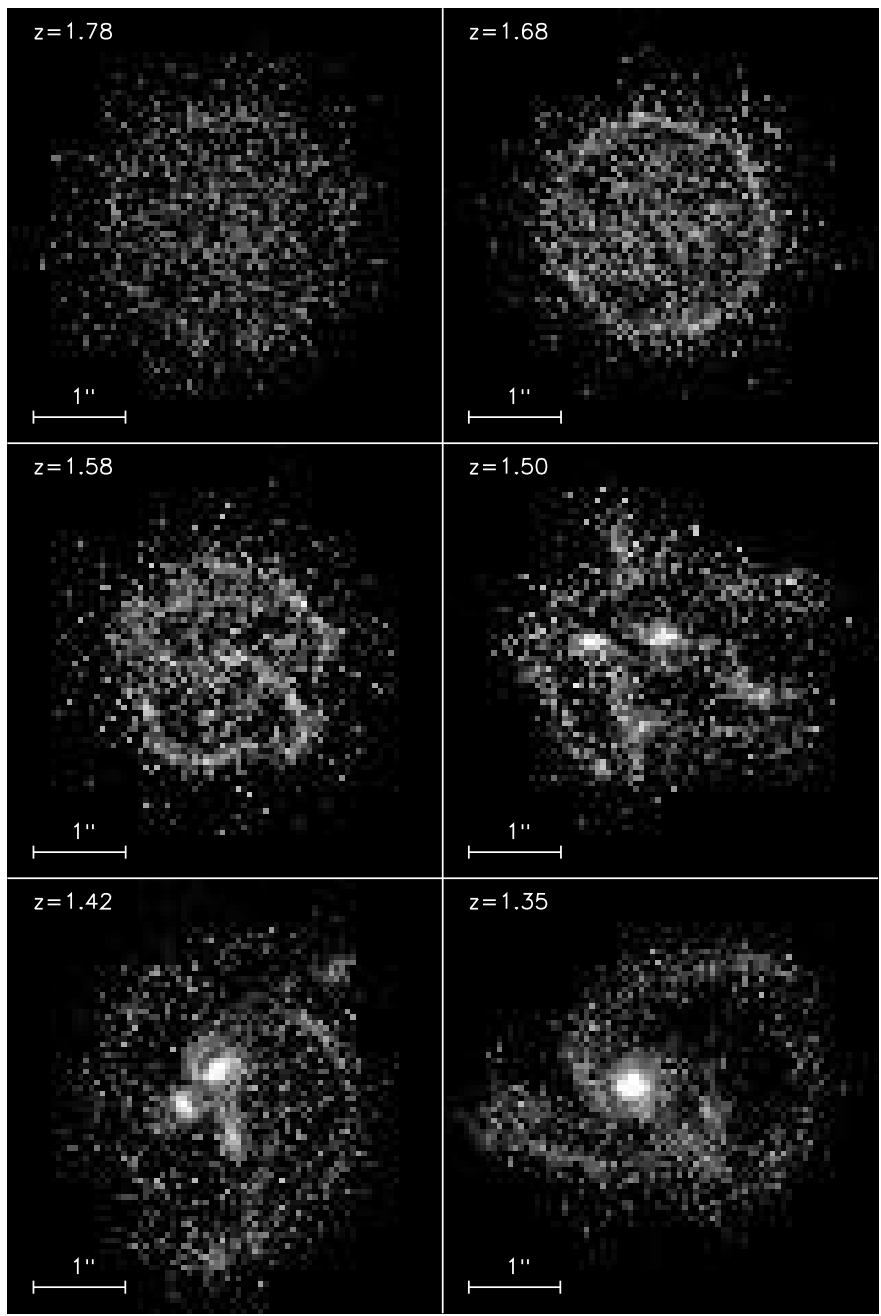}
         \caption{
                 Fragmentation phase of the star-forming disk model,
                 shown in observed F606W surface brightness, starting 
                 at 0.5~Gyr at top left and continuing in
                 200~Myr intervals. For K-correction and surface
                 brightness dimming the middle right panel was shifted
                 to  $z=1.5$, a typical redshift for chain galaxies.
                 Redshifts of other panels are relative to the $z=1.5$
                 panel and are indicated in each
                 map.  The frames are 40~kpc a side. HST resolution and
                 a detection limit of 28.21 mag \citep{williams96} were
                 used. Angular diameters were calculated using a
                 $\Lambda$CDM cosmology with $\Omega_M = 0.3$,
                 $\Omega_\Lambda = 0.7$, $h=0.7$.}
         \label{sequenz}         
         \epsscale{1.0}                                                         
         \end{figure}

However, at around 700 Myr the gas disk becomes unstable on large
scales and begins to fragment. The lower-mass stellar system follows
the gravitational potential perturbations induced by the gas.
Fig.~\ref{sequenz} shows a face-on view of the model evolution from
500-1500~Myr in terms of observed HST F606W surface brightness.  The
model was shifted to the redshift range indicated in the frames.

To quantify the fragmentation we have used the asymmetry parameter $A$
\citep{abraham96} in rest frame U-band.  The evolution of $A$ is very
similar to that of the SFR, which illustrates that the SF is driven by
fragmentation.  The high symmetry in the first two images reflects the
symmetric infall of the gas. The pressure from the SF in the disk and
the pressure from the infalling material causes the development of the
ring-like structure at the border of the stellar disk, visible in the
second image. This ring structure represents less than $10\%$ of the
total mass of the cloudy medium in the disk at this time. Yet the
enhanced SF in this structure, due to feedback-induced large density
fluctuations, causes a very prominent UV emission shifted to F606W at
the redshift considered.

\begin{figure}
         \epsscale{0.7}                                                         
         \plotone{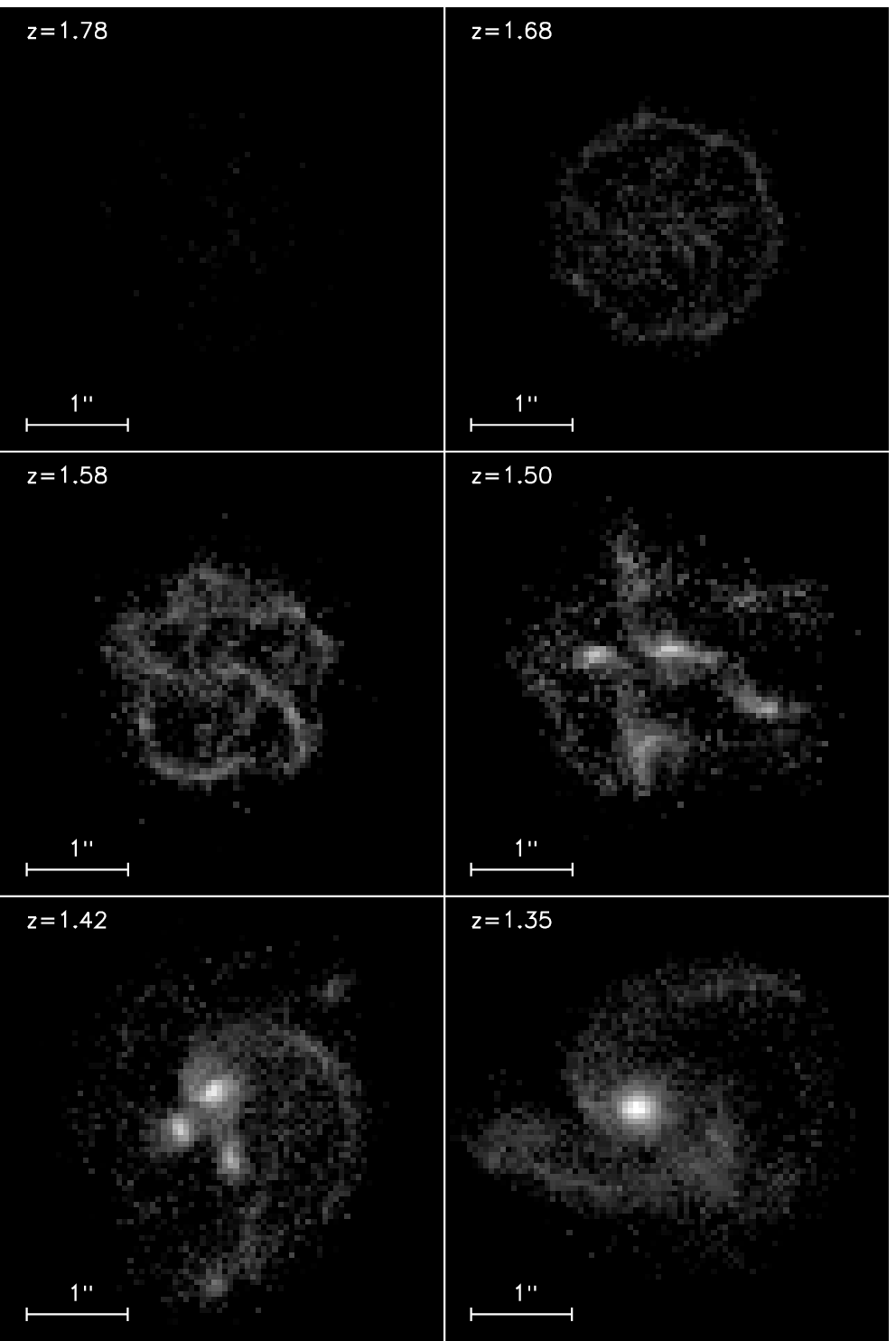}
         \caption{
                 Fragmentation phase of star-forming disk model, as in 
                 Fig.~\ref{sequenz}, but in NICMOS H160 surface
                 brightness. Here we used a limiting surface brightness
                 of 25.05, corresponding to the
                 $10\sigma$ limit of 26.1 mag in an aperture of 0.''7
                 diameter in the observations of \citet{ellis01}.
                 Notice that the clumpy disk structure is also 
                 visible in infrared passbands, showing that not only
                 the light distribution, but also the mass
                 distribution of the disk is fragmented.}
         \label{sequenzk}         
         \epsscale{1.0}                                                         
         \end{figure}

It is important to note that the disk shows its clumpy structure also
in the H-band (Fig.~\ref{sequenzk}), which would be traced by NICMOS
observations. This emphasizes the fact that the clumps are not only
regions of high star formation in an underlying smooth disk, but that
the disk itself is fragmented, forming stars vigorously in several
dense clumps. This makes it hard to distinguish such clumpy disks from
a merger event on the basis of observed surface brightness maps alone.

The maximum SFR is reached in this simulation during the main
fragmentation phase, after about 1 Gyr.  The morphology of the disk at
this time is bracketed by panels 3 and 4 of Fig.~\ref{sequenz}. The
SFR at this time is around 220~$\msun yr^{-1}$, corresponding to a
strong starburst galaxy.
Indeed, many high-redshift objects may be starburst galaxies.
\citet{lowenthal97} report similarities in stellar emission and
interstellar absorption lines between $z\sim 3$ galaxies and local
starburst galaxies.  The color selection criteria for Lyman Break
Galaxies (LBG) also strongly favor starburst galaxies \citep{steidel96}.
Age determinations of stellar populations and enhanced abundances of
$\alpha$-elements in LBGs \citep{carollo01} indicate that the 
very high SFRs in these high-redshift objects
persist only for a few hundred Myrs.

The clumps that form in the disk during the fragmentation phase spiral
to the center, building a massive bulge (last two panels of
Fig.~\ref{sequenz}). The clumps lose their angular momentum by
dynamical friction against the massive arms they generate in the disk.
Because a substantial fraction of the mass in the region of interest
by then consists of baryonic material, the timescale for this
spiral-in phase is relatively short, of the order of two disk rotation
times. This is typical for all models with a fragmenting gas disk
discussed in \citet{immeli04}.

The strongly asymmetric potential leads to a redistribution of angular
momentum, which is partially carried away by stars which leave the
simulation volume.  The high mass of the bulge is explained by the
efficiency of the angular momentum transfer during the fragmentation
phase.

Without newly infalling material there will be no major changes in the
global structures of the galaxy after 2.5~Gyr.  Formation of a bar is
prevented due to the high-mass bulge.  Numerical investigations of
\citet{noguchi99} showed a qualitatively similar evolution of a
gaseous disk. Two main differences in his work are that the
dissipative evolution is described by only a single sticky particle
phase, and that in his model the high SF threshold allows SF only
after fragmentation of the disk.

The stability properties of multi-phase star-gas disks such as the
model analyzed here, and the implications for bar and bulge formation,
are discussed in more detail in \citet{immeli04}.

\subsection{Comparison with Observations}

\begin{figure}
        \epsscale{0.7}                                                         
        \plotone{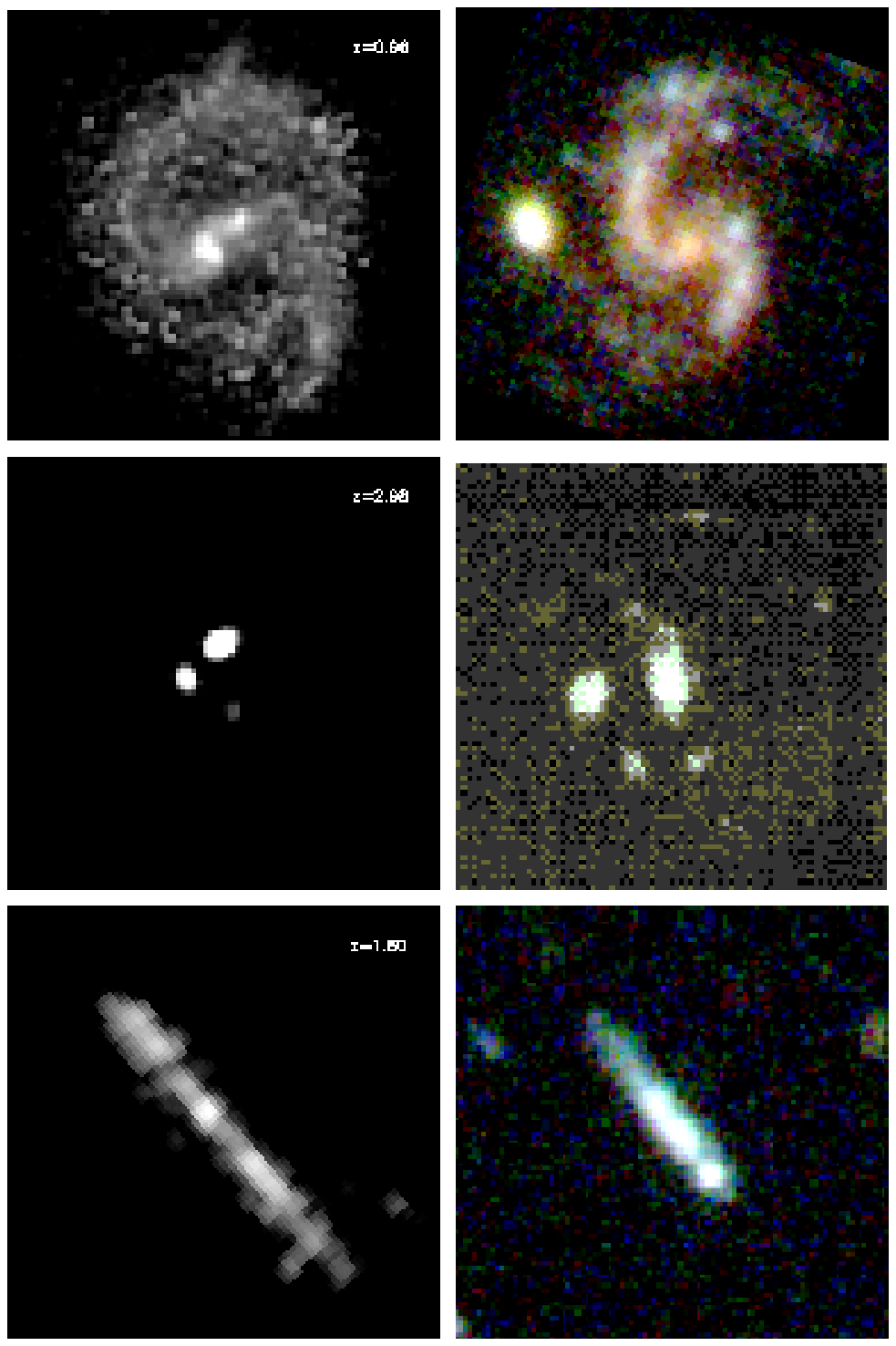}
        \caption{Comparison of observed HST F606W surface brightness 
                 of the star-forming disk model (left panels) 
                 with observations from vdB96 (right panels). 
                 For this comparison the model was shifted to the indicated 
                 photometric redshifts of the observed galaxies
                 \citep{fernandezsoto99}.}
        \label{comp}
        \epsscale{1.0}                                                         
        \end{figure}

In Fig.~\ref{comp} the model morphology at different times during the 
fragmentation phase is compared to the morphologies of some objects in the HDF. 
Clearly, several HDF morphologies can be well explained by the fragmented 
disk model.

%
%
%

In the first row of Fig.~\ref{comp} we show a phase of enhanced spiral 
arms at about 1.35~Gyr in the model, induced by the merging of the last 
two massive clumps, and a similar object observed in the HDF.  
In the second row we compare the model at 1.3 Gyr to a clumpy structure
at high redshift.  Seen edge-on (third row), the model resembles a chain 
galaxy during its fragmentation phase.

%
CHS95 reported observations of chain galaxies (chains), a new population of
high redshift galaxies observed with HST in the Hawaii Survey Fields,
with high major-to-minor axis ratios and very blue colors.  vdB96 also
found chain galaxies in the HDF.  

As already mentioned in the introduction, there are different
explanation scenarios for the chain galaxies. Recent observations
point to the fact that chain galaxies are indeed edge-on disks
\citep{elmegreen04a}. Comparison of the clump colors with the colors
of clumpy face on disks (sometimes called clump clusters) leads to
similar results, indicating that these clump clusters are indeed
face-on counterparts of chain galaxies.  Additionally these authors
found that the distribution of axial ratios for chain galaxies and
clumpy disks is similar to the distribution of local disk galaxies.
        \begin{figure}
        \plotone{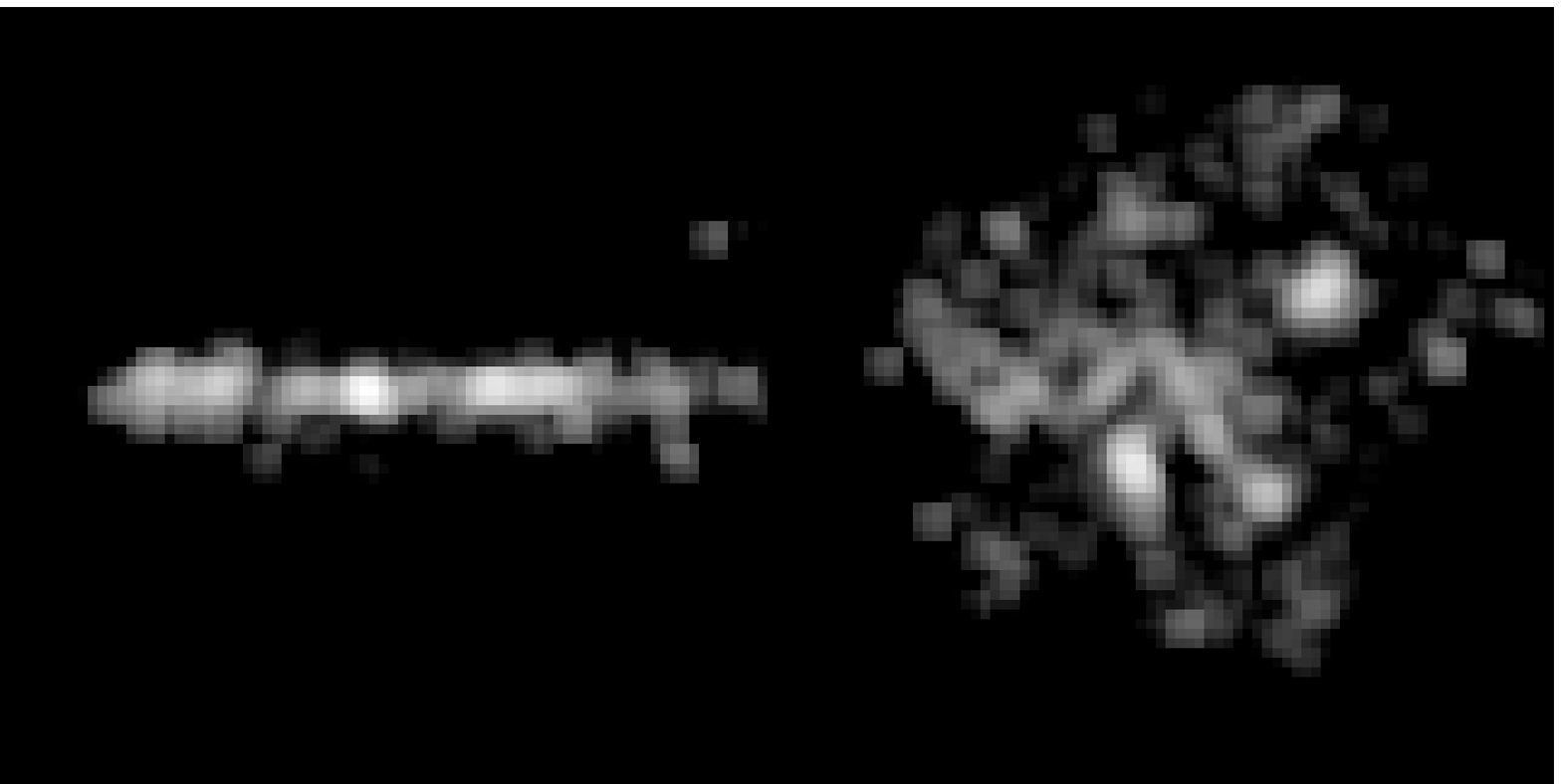}
        \caption{The chain galaxy model (at time 1.15~Gyr, shifted to
                 $z=1.6$, as in Fig.~\ref{comp}), edge-on and face-on in 
                 observed F606W surface brightness. Seen face-on, the chain
                 model resembles the clumpy disks observed by 
                 \citet{elmegreen04a}.}
        \label{chain_clump}
        \end{figure}

\begin{figure}
        \plotone{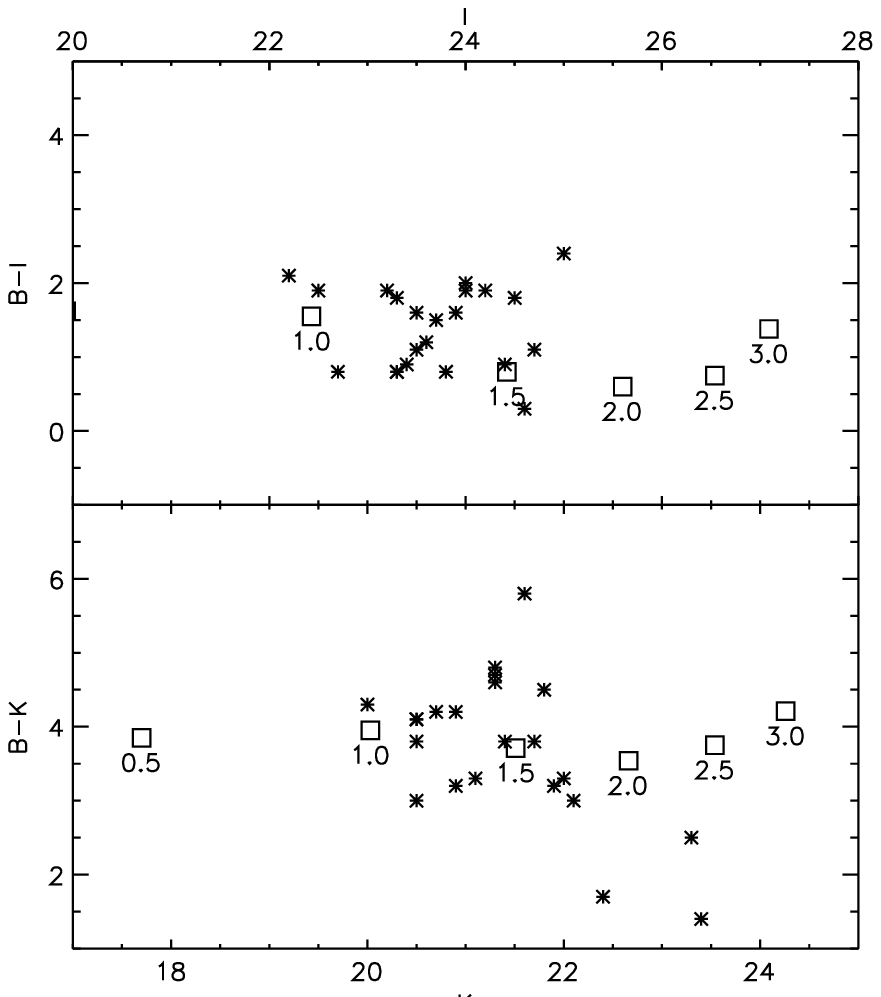}
        \caption{Colors of the model seen edge-on at 1.15~Gyr,  shifted to the
                 indicated redshifts (squares).  Observational 
                 data for chains from CHS95 (stars).}
        \label{cowievgl}
        \end{figure}

Given the small number of chains compared to approximately 1000 known
high redshift galaxies \citep{giavalisco02}, it is likely that these
objects are in a short evolutionary phase.  Our clumpy disk
model indicates that
the interpretation of \citet{oneil00} and \citet{elmegreen04a} is
correct.  It shows chain structures when viewed edge-on and during a
period of very high SFR; see Fig.~\ref{chain_clump}.  
Because this period is short compared to a
Hubble time, these objects will be relatively rare, and because of the
high SFR, they are very blue.  The model therefore naturally explains
also the observation of CHS95 that a large fraction of chains is very
blue. A comparison of the model colors with those of CHS95 is shown in
Fig.~\ref{cowievgl}.  Best agreement is obtained if our model is
shifted to redshifts between 0.8 and 1.8 (see also \citet{immeli03}).

\begin{figure}
        \epsscale{0.7}                                                         
        \plotone{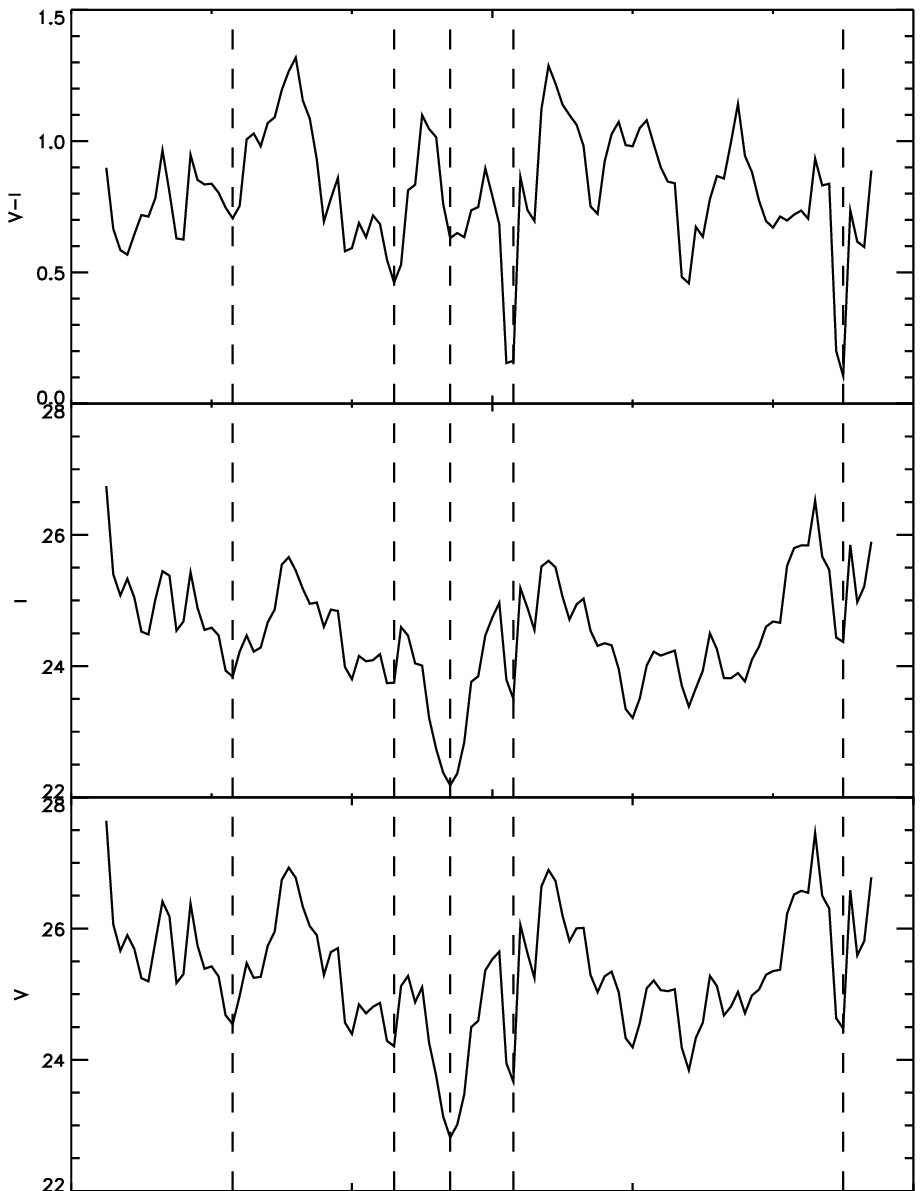}
        \caption{Color profile of the clumpy disk model seen edge-on,
                 after 1.15~Gyr, when shifted to $z=1.6$ as in Fig.~\ref{comp}.
                 The positions of the main clumps are indicated by the vertical
                 dashed lines. The figure shows that
                 the typical flat profile for chain galaxies 
                 \citep{elmegreen04b} is nicely reproduced, and that the
                 clumps as regions of intense star formation are both
                 bright and blue in V-I color, again as observed.}
        \label{elmegreenvgl}
        \epsscale{1}                                                           
        \end{figure}

Comparing the color profile of the chain galaxies with observations
(Fig.~\ref{elmegreenvgl}) shows also good agreement.  The model
reproduces the flat profile, which in our model is a direct
consequence of the constant-surface density infall.  The typical
exponential profile observed in present-day disk galaxies only emerges
in the instability phase. The scale length of the resulting
exponential disk depends on the infall radius.

Because of the strong instabilities in the disk, gas and stars are
dynamically heated. The gas, on the other hand, is also cooled
by dissipative cloud collisions. In the present model,
the velocity dispersion of the gas takes values between 30 and 50~$\kms$, 
about 25\% of the maximal rotation velocity. This is consistent with 
estimates from observations \citep{elmegreen04a}. Also the number of 
clumps and their masses are in perfect agreement with these observations.


\subsection{Local galaxies}

Do gas-rich disk systems at low redshift also tend to form fragments
with enhanced SF? An example of a nearby galaxy in which the
fragmentation process may be taking place, is the gas-rich, blue
starburst galaxy NGC 7673 \citep{homeier99}. This object has a
remarkably clumpy morphological appearance, seen in both the R-band
and H$\alpha$, even though the H$\alpha$ velocity field is that of a
regular, rotating disk.

Dwarf irregular galaxies observed in the local universe generally have
a high gas fraction and often a disk-like structure.  Recent
observations \citep{billett02} show that there is a tendency for star
formation to be concentrated in localized regions of high column
density. Due to the lower mass of the dwarf galaxies, one cannot
directly compare them to our model. In particular, the dynamical
influence of the dark halo is likely to be more important in
dwarf galaxies; explicit models of such lower mass galaxies are
needed.


\section{Merger or Fragmentation?}

\begin{figure}
        \plotone{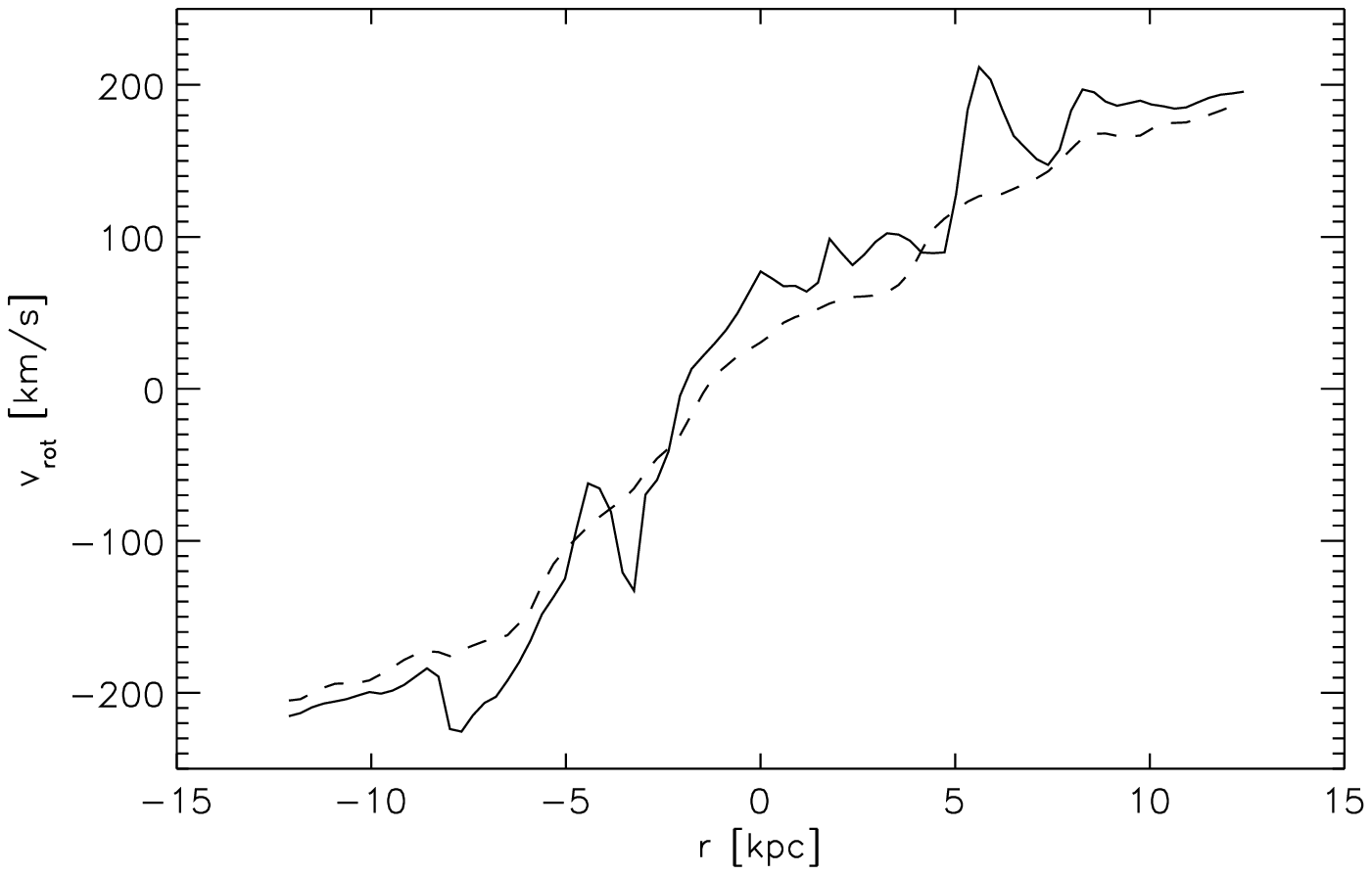}
        \caption{Mass-weighted rotation curve of the gas after 1.15~Gyr,
                 in the disk plane where the clumps dominate (solid line), and 
                 0.5~kpc above the plane (dashed line). The deviations from
                 the smooth rotation curve are significantly larger than 
                 expected purely from the velocity dispersion in the disk.
                 They reflect the gravity of the massive clumps in the disk.}
        \label{rotchain}
        \end{figure}

Kinematical data will be important to further clarify the nature of
chain galaxies.  While in our fragmented disk model the massive clumps
should still show the underlying disk rotation, in a merger scenario
no similar alignment of the clump velocities is expected.
Fig.~\ref{rotchain} shows the predicted influence of the clumps on the
rotation curve of the gaseous disk; deviations from the smooth
rotation profile are up to 100~$\kms$.  These deviations are due to
the gravitational influence of the massive clumps; they are much
larger than expected from the velocity dispersion in the disk alone.
In summary, while the basic rotation pattern remains visible during
the fragmentation phase, it is severely disturbed by the innermost
brightest knots.

\begin{figure} \centering
        \epsscale{0.8} 
        \plotone{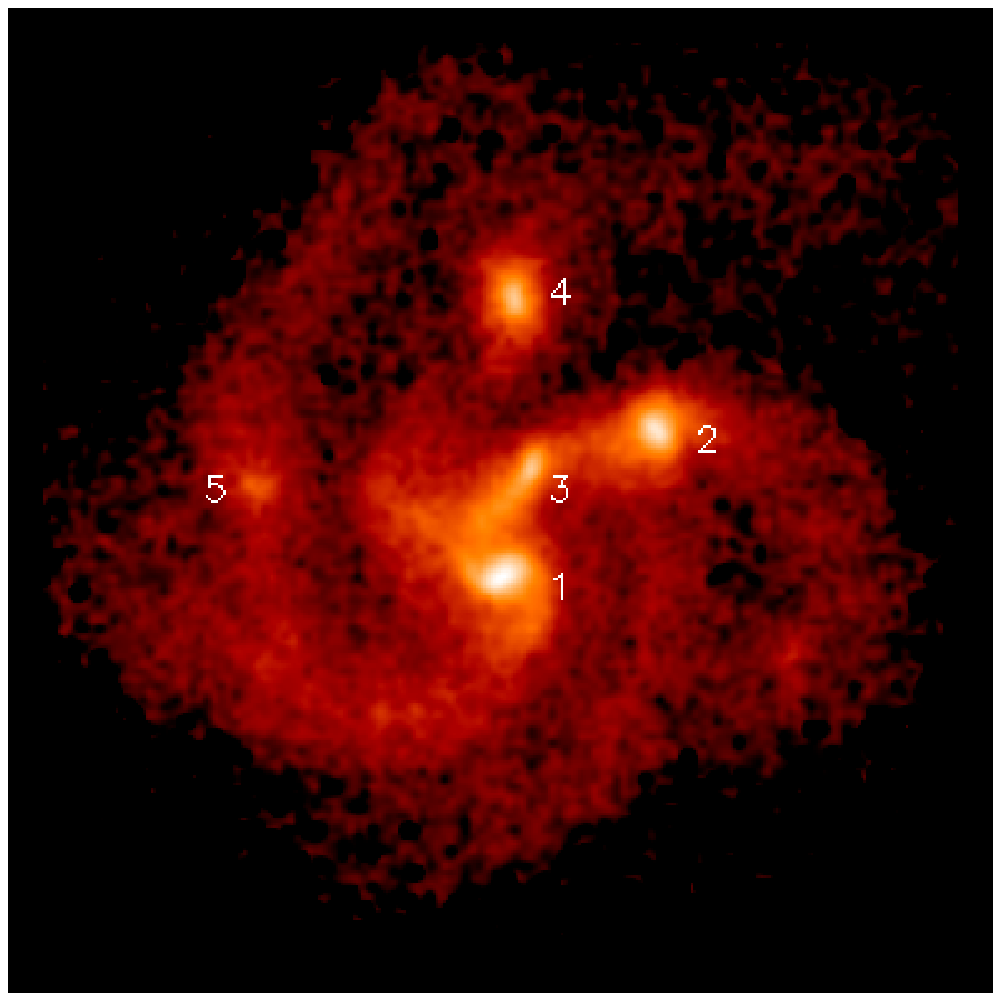}
        \caption{Smoothed stellar mass surface density of the model 
                 at 1.2~Gyr with numbering of the clumps.}
        \label{klumpen}
        \epsscale{1.0} 
        \end{figure}

%
Additionally, the metallicities of the clumps in a merger event are
expected to vary significantly, depending on the mass and evolution
history of the merging clumps. Contrarywise, one expects similar
metallicities for the clumps in a fragmented disk.  We investigate the
metallicity of five clumps selected in the fragmented disk model as
indicated in Fig.~\ref{klumpen}.  We get abundance differences of up
to 0.25 dex (Table \ref{datatab}), depending on the masses of the
clumps ($\sim {\rm few}\, 10^9\, \msun$ for those in Fig.~\ref{klumpen}).
No differences in oxygen-to-iron ratios can be measured, due to
dominance and young age of the starburst.

%

%
%
Many authors report observations of multiple knots \citep[][
vdB96]{driver95, steidel96}.  The fragmentation scenario naturally
explains objects consisting of several clumps, whereas in a merging
scenario it is much less likely to see more than three nearby clumps
merging at the same time.  Also, the synchronized colors often
observed in these clumpy objects \citep{abraham01} are naturally
explained with the fragmented disk scenario. In the present model, the
mean age of the stars seen edge-on in the clumps is constant within 70
Myr.  At least some of the observed  multi-clump systems may therefore
represent fragmented disks. This is confirmed by the recent observations 
of \citet{elmegreen04a}: a merger event would lead to more 
spheroidal systems than the thin chain galaxies.


\section{Conclusions}
A short formation timescale of a galactic disk due to a high
dissipation rate for the cold gas phase leads to fragmentation and to
the formation of a clumpy disk with an enhanced SFR in the clumps.
After the fragmentation phase the clumps fall to the center building a
massive bulge.  Subsequent bar formation is prevented by the massive
bulge.

Chain and multi-clump morphological structures, as well as
synchronized colors observed in high redshift objects, can be well
explained by a fragmented disk in a gas rich, single galaxy.  Chain,
double and tadpole galaxies may be different evolutionary states of a
fragmented disk.  Our model suggests that these galaxies are in their
formation process and are observed during their relatively short
fragmentation phase, with a high SFR, comparable to the model SFR of
up to $220~\msun~yr^{-1}$.  This high SFR is generated by a disk
instability alone, and there is no need for external triggering
through interactions or a merger with other galaxies.

The effects of single clumps on the mass-weighted rotation curve
in our model can be as high as 100~$\kms$. Nonetheless the underlying
rotation signature should be observable.  Metallicity differences in
the clumps of the fragmented disk are no larger than 0.25 dex,
while their mean stellar ages are highly synchronized.
Observations to test these predictions are highly desirable.

\vspace{0.5cm}
We thank the Schweizerischen Nationalfonds for financial support of this
work and the Centro Svizzero di Calcolo Scientifico (CSCS) for giving us the
opportunity to use their computing facilities.

\clearpage

        \begin{table}[h]
        \begin{center}
        \begin{tabular}{ccccc} 
        \tableline
        \tableline
        Clump       & Mass [$10^9 \msun$]  & [O/H] & [Fe/H] & [O/Fe] \\ 
        \tableline
        1&      7.96&     -0.32&     -0.72&      0.40\\
        2&      4.10&     -0.38&     -0.77&      0.39\\
        3&      2.02&     -0.44&     -0.84&      0.40\\
        4&      2.39&     -0.42&     -0.82&      0.40\\
        5&      0.64&     -0.59&     -0.98&      0.39\\
        \tableline
        \end{tabular}
        \caption{Masses and metallicities of the stars in the clumps
                 at 1.2~Gyr (see Fig.~\ref{klumpen}). Metallicity
                 differences are up to 0.25 dex. [O/Fe] is constant
                 because SNe Ia with a typical time delay of around
                 1~Gyr have not yet contributed significantly to the
                 chemical abundances.}
        \label{datatab}
        \end{center}
        \end{table}

\clearpage


\end{document}